\documentstyle[prl,aps,epsf,floats,twocolumn]{revtex}

\begin{document}


\wideabs{

\title{Novel dynamical effects and persistent memory in phase
separated manganites}
\author{P. Levy, F. Parisi, L. Granja, E. Indelicato and G. Polla.}
\address{Departamento de F\'\i sica, Comisi\'{o}n Nacional de
Energ\'\i a At\'{o}mica,Gral Paz 1499 (1650) San Mart\'\i n,Buenos
Aires,Argentina}
\date{29 October 2001}
\maketitle

\begin{abstract}

The time dependent response of the magnetic and transport properties of
Fe-doped phase separated (PS) manganite La$_{0.5}$Ca$_{0.5}$MnO$_{3}$ is reported.
The nontrivial coexistence of ferromagnetic (FM) and non
FM regions induces a slow dynamics which leads to time relaxation and cooling rate
dependence within the PS regime. This dynamics influences drastically on physical
properties. On one hand, metallic like behavior, assumed to be a fingerprint
of percolation, can  be also observed before the FM phase percolates as a result
of dynamical contributions. On the other one, two novel effects for the manganites
are reported, namely the rejuvenation of the resistivity after ageing, and a
persistent memory of low magnetic fields ($<$ 1 T), imprinted in the amount of the FM phase.
As a distinctive fact, this memory can be recovered through transport measurements.

\end{abstract}

}

\narrowtext

Phase separation (PS), namely the simultaneous
presence of submicrometer ferromagnetic (FM) and charge ordered (CO)
regions, is emerging as the most important issue in the
physics of the manganese-oxide-based compounds. \cite{Dagotto}
The PS scenario appears as  particularly favorable for the existence
of out-of equilibrium features. The competition between the
coexisting phases opens  the possibility for the appearance of
locally metastable states, giving rise to interesting time
dependent effects, as
cooling rate dependence, \cite{Uehara2} relaxation,
\cite{Voloshin,Babushkina,Smolyaninova} giant $1/f$ noise,
\cite{Podzorov1/fnoise} two level fluctuations, \cite{TLF} non
equilibrium fluctuations, \cite{PodzorovFluct} noise \cite{Anane} and
relaxor ferroelectric like behavior. \cite{Kimura} The
similarity between PS manganites and glassy systems, coming from the
frustration of the FM and the CO states at the phases boundary was
also suggested. \cite{RoyJAP} 

Most of the relaxation experiments were performed after the induction
of metastable states by, for instance, application of  magnetic
fields $H$ after zero field cooling, \cite{Anane,Kimura,RoyJAP}
removal of $H$ after field cooling, \cite{Smolyaninova} x-ray
ilumination, \cite{CasaEulett} electron beam irradiation,
\cite{PodzorovFluct} etc. Such external perturbation can also be an
abrupt change of the temperature, as observed in the prototypical
PS compounds La$_{5/8-y}$Pr$_y$Ca$_{3/8}$MnO$_3$ ($y=0.35$) and
La$_{0.5}$Ca$_{0.5}$MnO$_{3}$.
\cite{Uehara2} The extremely slow
relaxation observed in all the mentioned works opens an interesting
question not previously addressed about the path followed towards
equilibrium in the PS state of manganites. 

In this work we present a detailed study of time dependent effects in
the PS compound La$_{0.5}$Ca$_{0.5}$Mn$_{0.95}$ Fe$_{0.05}$O$_{3}$
(LCMFO). The parent compound La$_{0.5}$Ca$_{0.5}$MnO$_{3}$ exhibits PS
with the temperature of charge
ordering $T_{co}$ lower than that of the ferromagnetic order
$T_C$. \cite{LaCa} From a microscopic point of view, the inclusion of Fe in the Mn site of
La$_{0.5}$Ca$_{0.5}$MnO$_{3}$ yields the same overwhelming effect on
the CO state as Cr doping does on other half doped manganites as Nd$_{0.5}$Ca$_{0.5}$MnO$_{3}$
\cite{Kimura}
and Pr$_{0.5}$Ca$_{0.5}$MnO$_{3}$, \cite{MahendiranCr}
a feature ascribed to the presence of a
random field quenched by the impurities. \cite{Katsafuji}
But, unlike what happens with Cr doping, Fe doping also disrupts the double
exchange interaction due to its filled e$_g$ orbitals. Moreover, Fe-O-Mn
superexchange interactions are likely to be antiferromagnetic \cite{Morrish}.
All these facts are reflected by the decrease of $T_C$ as a function of Fe
doping. \cite{Recife}

We use transport and magnetization measurements to study the dynamic
of the coexisting phases, which is observed close below T$_C\approx$
90 K. We found cooling rate dependence and slow relaxation effects in
the temperature range between T$_C$ and 50 K, indicating the
enlargement of the FM phase as the PS state evolves with time. Within
this scenario, our main results are two novel effects. One, the rejuvenation in the resistivity
curve when
cooling is resumed after ageing, which resembles the behavior of
glassy systems \cite{Nordblad,Bouchaud} and disorder ferromagnets.
\cite{Vincent} The other is the effect produced by the application of
a low $H<$ 1 T while ageing which, instead of inducing metastable
states, can drive the system towards its zero field equilibrium
point. As this process is irreversible, the effects of $H$ on the
resistivity remain even after it is removed, and the system keeps
memory of the magnetic history, acting as a magnetic field
recorder.

Polycrystalline samples of LCMFO were synthesized by the sol-gel
technique, their average grain size was around 0.5 microns. DC resistivity was
 measured using the four probe
technique, magnetization was measured using a commercial SQUID
magnetometer. Figure 1 displays magnetization $M$ and
resistivity $\rho$ as a function of temperature on cooling. $M$(T)
increases continuously between 100 and 50 K, reaching a low
temperature plateau which reveals a mostly FM state. $M$ vs $H$ loops
at 5 K show a saturation magnetization of $\approx$ 3 $\mu_B$/Mn at 5
T.  An insulator to metal transition is suggested at T$_p\approx$ 80
K. Both $T_p$ and the resistivity below $T_D\approx$ 85 K were found to
be very sensitive to the cooling rate $v_c$ (inset Fig. 1). The dependence of
the peak resistivity with $v_c$ reassures that this cooling rate effect is not an experimental
artifact.

We studied the time relaxation of $\rho$ at several $T < T_D$
using  $v_c$=0.2 K/min as the
cooling rate to approach and depart from the values at which $T$ was
stabilized for one hour (Fig. 2). The relaxations are characterized
by the decrease of $\rho$
following a logarithmic time dependence, and are as high as an 18\%
in one hour at 73.8 K. A related behavior (inset of Fig. 2) was
observed in the relaxation of $M$. Both relaxations are consistent
with the isothermal growth of FM regions embedded in a non FM host.
\begin{figure}[t] 
\centering
\epsfysize=6.0 cm
\centerline{\epsffile{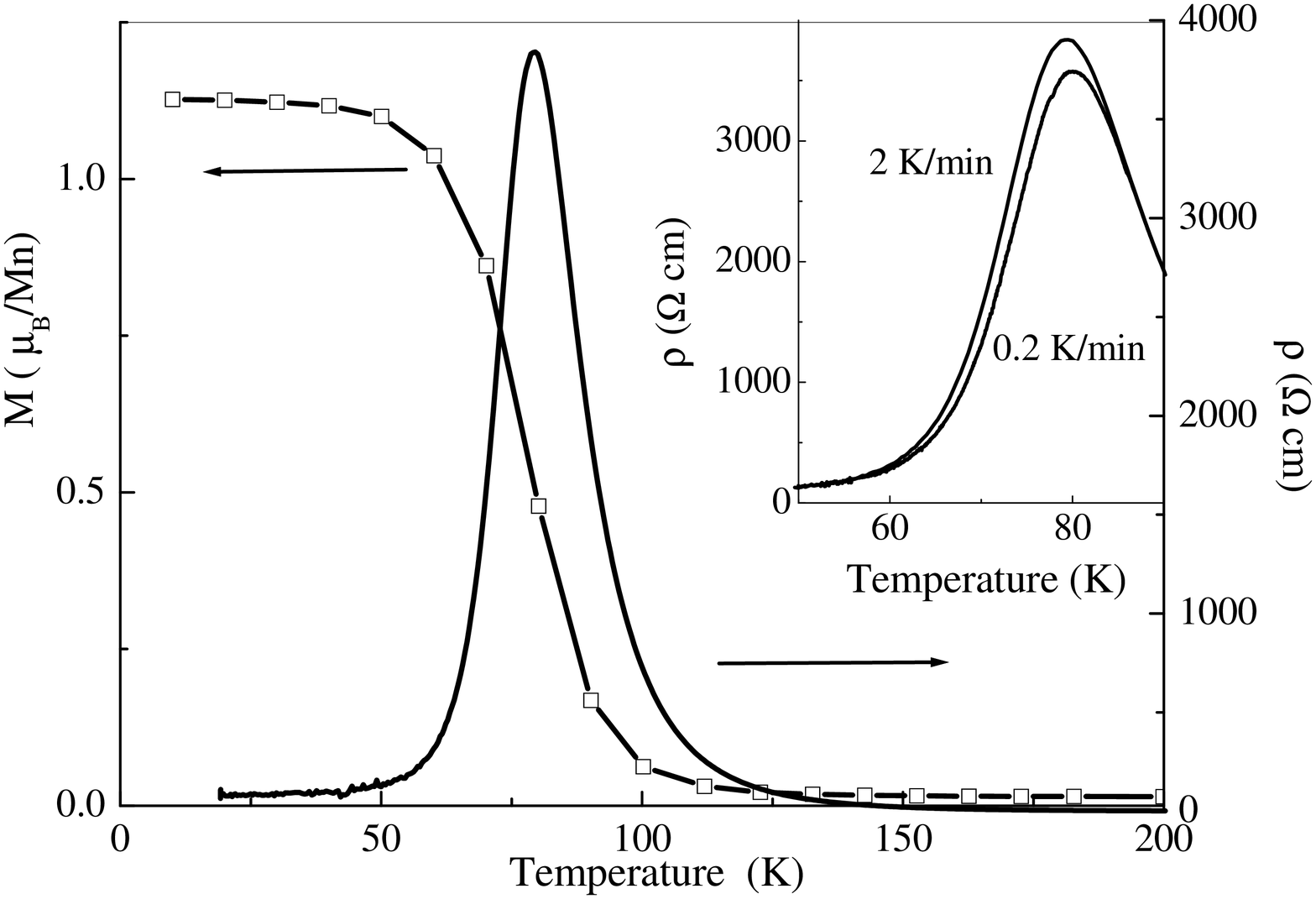}}
\caption{Temperature dependence of $M$ ($H$ = 0.1 T) and $\rho$ ($H$
= 0) on cooling for LCMFO. Inset: $\rho(T)$ when cooling at $v_c$= 2
and 0.2 K/min. 
\label{Fig1}
}
\end{figure}
A noticeable fact is that, when the cooling process is resumed with
$v_c$= 0.2 K/min after relaxation, $\rho$ merges smoothly with the
curve obtained at that $v_c$ without $T$ stabilization steps. This
effect is similar to that found in the dissipation of some disordered
ferromagnets and spin glasses, \cite{Nordblad,Bouchaud,Vincent} and
has been named rejuvenation. Another striking feature is that the
slope of the $\rho (T)$ curve in this range is highly dependent on
the previous history of the sample. For instance, while
$\partial\rho/\partial T > 0$ is found when cooling continuously,
suggesting metallic behavior, a typical insulator response is
obtained if, after one hour ageing, cooling is resumed (Fig. 2).

This last result is opposite to that expected from a static vision of
transport properties in PS systems, in which the change from
metallic-like to insulator-like behavior is driven by a decrease in
the amount of the metallic fraction. This fact points to the
existence of dynamical contributions to the resistivity in the
continuously cooling procedure. To get a better description of  this
behavior we have studied  the time dependence of the resistivity
slope $\partial\rho/\partial T$ by performing small thermal cycles
($\Delta T\approx$ 0.6 K) around a fixed $T$ value while the system
is relaxing (Fig. 3). After a sudden initial change from positive to
negative, the slope $\partial\rho/\partial T$ increases smoothly and
slowly (see Fig. 3 inset) as the system evolves. The obtained ageing
behavior of $\partial\rho/\partial T$ is now consistent with the
static image for transport, the negative slope indicating that
percolation of the FM phase has not been achieved close below $T_p$.
As in this $T$ range the state of the system is characterized by the
slow enlargement of the FM regions against the non FM host, the
approach to equilibrium drives the system towards the percolation
threshold, increasing the resistivity slope. The metallic-like
behavior of the continuously cooling curve is then
to be ascribed to the existence of non-static contributions to
$\partial\rho/\partial T$ arising from the out-of-equilibrium
dynamics of the coexisting phases.
As the approach to equilibrium becomes slower as time goes by, it is
not possible to reach it within laboratory times, so at this point we
can not give conclusive statements about the nature of the
equilibrium state (homogeneous FM or PS). However, as the system
evolves by increasing the FM phase, an external $H$ may assist in the
path towards equilibrium. Following this idea, we have studied the
effect of low $H$ applied for a short time while the system is
relaxing after zero field cooling the sample to a  T close
below $T_p$. 
\begin{figure}[t] 
\centering
\epsfysize=6.0 cm
\epsffile{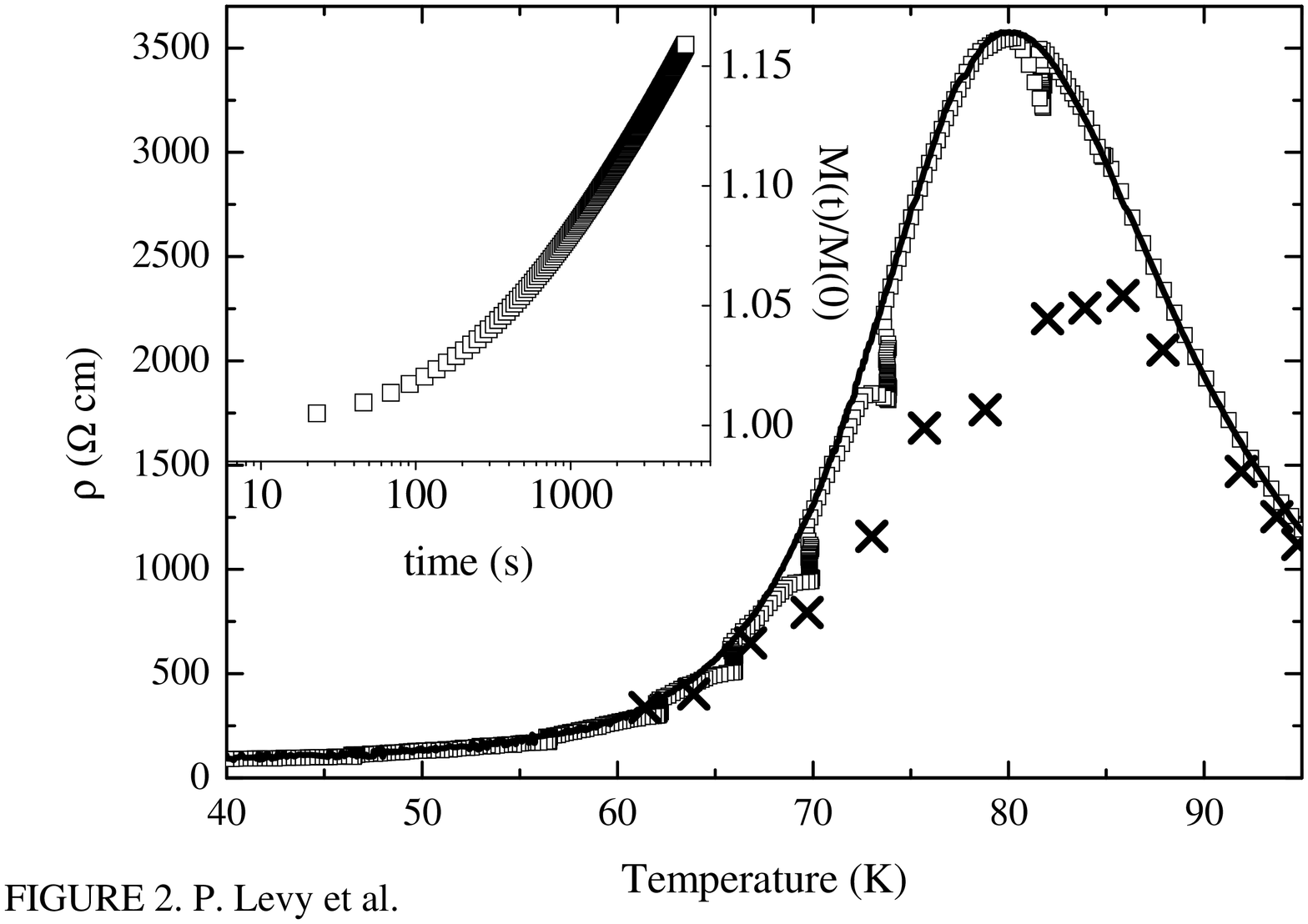}
\caption{ Temperature dependence of $\rho$ on cooling at 0.2 K/min
continuously (solid line) and with intermediate stops of one hour
(open symbols). The crosses indicate the $\rho$ values corresponding to the equilibrium state.
Inset: time dependence of $M$ after zero field
cooling to 80 K and application of $H$ = 0.1 T. 
\label{Fig2}
}
\end{figure}
              
\begin{figure}[t] 
\centering
\epsfysize=6.0 cm
\epsffile{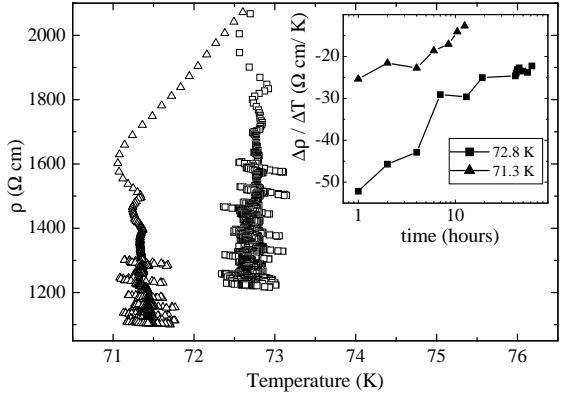}
\caption{ Resistivity relaxation data obtained at 71.3 and 72.8 K;
while the system was ageing, small thermal cycles ($\Delta T\approx$
0.6 K)
were performed at 0.2 K/min, to evaluate $ \partial
\rho/\partial T$. After these sweeps, cooling was resumed at 0.2 K/min
 confirming that the data rejoins the dynamic curve. Inset: time dependence of $\partial
\rho/\partial
T$ at 71.3 and 72.8 K.
\label{Fig3}
}
\end{figure}  
Fig. 4-a displays $\rho$ as a function of elapsed time upon the
application of several $H$ values at $T$= 72 K, showing jumps when
the field was applied and removed. The sudden decrease of $\rho$ when
the field is turned on can be ascribed to two independent mechanisms,
one originated in the alignment of spins and domains, and the other due to the
enlargement of the FM phase driven by $H$. \cite{Parisi} It is worth noting the hysteretic
behavior of $\rho$ after application and removal of $H$, indicating that the system keeps memory
of the magnetic history imprinted in its zero field resistivity.

The slow relaxation of the persistent resistivity value observed
after $H$ was removed indicates that, depending on the strength of
the field, the FM phase may still be growing against the non-FM one
(Fig. 4-a inset). Upon increasing $H$, we found a $T$-dependent
threshold $H_{th}$ value above which the ulterior relaxation of
$\rho$ after the field is turned off reverses its
sign: after $H > H_{th}$ is applied and removed $\rho$ slowly
increases (instead of decreasing) indicating that the system is
evolving by diminishing the amount of the FM phase (Fig. 4-a inset).
This fact signs unambiguously that the equilibrium state is of PS
nature, characterized by the FM fraction $f_0(T)$. The
application of $H_{th}$ while ageing drives the system towards
its zero field equilibrium point, having an effect which is
equivalent to a long relaxation process. The equilibrium points
 depicted in Fig. 2 provide a demonstration of the magnitude of the dynamical
 effects. By applying a $H>H_{th}$ the
amount of the FM phase overcomes the equilibrium volume, leading to a
subsequent decrease of the FM fraction.

In what follows we discuss the overall results.
At T$_{C}$ an inhomogeneous FM state appears, consisting
in the coexistence of isolated FM clusters of definite size within a non-FM
host. At $T_D$= 85 K the FM
regions start to grow against the host material with the equilibrium
size of the clusters increasing  as $T$ is lowered. The process
followed by the clusters to reach their equilibrium size can be
thought as a stepwise movement of the phase boundaries through energy
barriers. Before the FM clusters reach the percolation threshold, the
temperature and rate dependent resistivity $\rho(T,v_c)$ can be
modeled as a series circuit, i.e.
\begin{figure}[t]
\epsfysize=10.0 cm
\epsffile{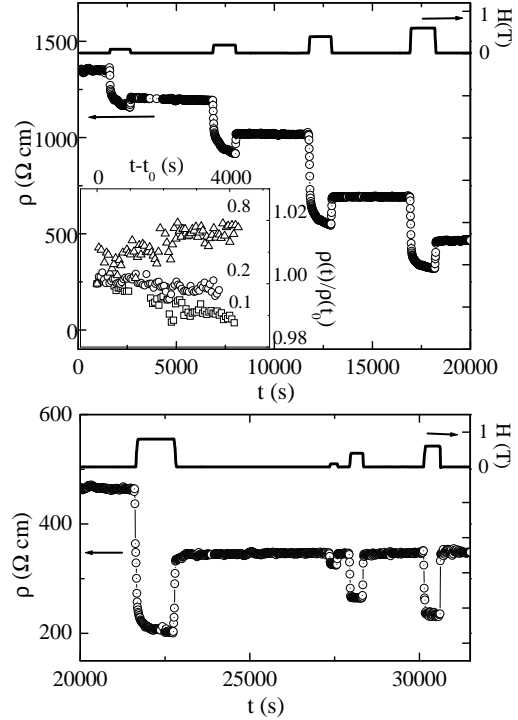}
\caption{ a) Resistivity (open circles) at T= 72 K as a function of time upon
application of $H$= 0.1, 0.2, 0.4 and 0.6 T (solid line).
Inset: time dependence of $\rho$(72 K) in $H$=0 normalized to the $\rho$ value
after the field $H$ was turned off; the labels are the $H$ fields. b)
Resistivity at T= 72 K as a function of time upon application of H =
0.1, 0.4 and 0.6 T after $H$= 0.8 T was applied and removed. 
\label{Fig4}
}
\end{figure} 
\begin{equation}
\rho(T,v_c)=n(T,v_c) \rho_{F}(T)+[1-n(T,v_c)] \rho_{nF}(T)
\end{equation}
where $\rho_{F}(T)$ and $\rho_{nF}(T)$ are the resistivities of the
FM and non-FM constitutive media respectively and $n(T,v_c)$ is a
measure of the relative fraction
$f(T,v_c)$ of the FM phase. As the system is rather close to the
percolation limit, $n(T,v_c)$ can be a cumbersome functional of
$f(T,v_c)- f_c$, (where $f_c$ is the percolation threshold), but is always a
monotonous increasing function of $f(T,v_c)$, and can be obtained
from conduction models through binary alloys.\cite{Mac} Within this
frame, the slope of the resistivity curve has two components. On one
hand, a static contribution given by
$n(T,v_c) \frac{\partial \rho_{F}}{\partial T}+(1-n(T,v_c))\frac{\partial
\rho_{nF}}{\partial T}$ 
which is typically less than zero in the non-percolative regime.
On the other hand, a "dynamical" term defined as
$\frac{\partial n(T,v_c)}{\partial T} (\rho_{F}-\rho_{nF})$
related to the change of
the size of the FM clusters as $T$ is varied. This term gives a
positive contribution to $\partial \rho/\partial T$ in the whole
range in which $f(T,v_c)$ increases as $T$ is lowered. As the system
is cooling down at a rate $v_c$, the appearance of energy barriers at
the clusters surface below $T_D$ prevent their free growth, and the
FM fraction $f(T,v_c)$ no longer follows its equilibrium value
$f_0(T)$, larger differences corresponding to higher $v_c$ values.
This fact accounts (Eq. 1) for the overall increase of $\rho$ as
$v_c$ is increased (Fig. 1, inset). Close below $T_D$, $f(T,v_c)$
increases slower than $f_0(T)$, yielding a low dynamical
contribution, i.e., larger(negative) slopes are achieved with higher
$v_c$. As the difference between $f_0(T)$ and $f(T,v_c)$ becomes
larger, the rate at which the FM clusters grow increases, and the
dynamical contribution approaches the static one. Both terms are of
the same magnitude at $T_p$, and below, the dynamical contribution is
 even larger than the static one. A positive slope resembling
"metallic-like" behavior is then obtained below $T_p$, although the
FM clusters do not percolate.

The history-dependent $\partial \rho/\partial T$ obtained after a
relaxation process (Figs. 2 and 3) is consistent with this
scenario. By ageing at a given temperature $f(T,v_c)$ slowly
approaches its equilibrium value $f_0(T)$ and the growth dynamics
becomes partially frozen. When cooling is resumed a small dynamical
contribution is obtained, compared with that of the non-stop process,
because of the "clamping" of the interface. The main contribution to
 $\partial \rho/\partial T$ after ageing comes then from
the static part, giving rise to the "insulator like" response. On
further cooling, the frozen-in state is released, the system falls
again in the dynamical regime mainly determined by $v_c$, and the
resistivity curves (with and without ageing) merge into a single one.

This dynamical process produces ageing and rejuvenation effects which bear similarities with
those found in glassy systems.
\cite{Nordblad,Bouchaud,Vincent} Slow relaxations following
stretched exponential or logarithmic dependences were accounted for
with models based on a hierarchical constrained dynamics,
\cite{Palmer} in which the system evolves through a hierarchy of
energy barriers, constrained to accomplish determined requisites
after a process at time $t$ can proceed. In our case the existence of
a hierarchy of energy barriers is revealed by the response of the
system to $H$ while ageing at a fixed $T$ showed in Fig.
4-a. When $H$ is applied all the barriers of height $< H$ are overcome,
yielding the sudden growth of the FM phase. As $H$ is increased,
higher barriers are crossed, giving rise to further enlargement
effect. In this context, once all the energy barriers of height up to
some applied $H_{MAX}$ were overcome, the subsequent application of a $H<H_{MAX}$ should have no
immediate effect on the relative fractions of the
coexisting phases. This picture is confirmed by the results shown in Fig 4-b. As
 can be seen, once $H_{MAX}= 0.6$ T has determined the
relative phase fractions, ulterior application of $H<0.6$ T produces only domain alignment,
without inducing additional changes on the amount of the coexisting phases. Then, after the
system was driven to a "close to equilibrium
state" by some $H_{MAX}$, the process can not be reversed and the system keeps memory of the
largest $H$ applied in its magnetic history.

The above described scenario seems to be characteristic of the low
T$_{C}$ PS systems. We have obtained very similar data in other
samples of LCMFO with slight different Fe doping and in
La$_{5/8-y}$Pr$_y$Ca$_{3/8}$MnO$_3$ ($y=0.30$), a PS compound with
 rather different hole doping but similar PS characteristics.

Summarizing, we have presented robust evidence of the importance of
the coexisting phase dynamics in the behavior of the PS manganites.
When dynamical effects are present the amount and spatial
distribution of the FM phase (percolative or not) can not be directly
inferred from metallic or insulator behavior. The rejuvenation found
in the resistivity curve after ageing and the persistent memory after
the application of a low $H$ are novel features in the physics of
manganites. The possibility to record an external magnetic field as a
sizeable and persistent change of the resistivity is a distinctive
fact of the memory effect, and opens a route for applications. Memory effects
 in manganites have been previously reported,
 \cite{Babushkina,Kimura,RoyJAP,CasaEulett}
related to the field induction of metastable states. In
our case, instead, the low field carries the system closer to its
 zero field equilibrium state. Since
both, the "virgin" state and the "H-enlarged" state are, in general, out of
equilibrium, the memory could be erased after a very long time, but
the effect is protected by the very slow dynamics displayed by the
system.

Project partially financed by CONICET, Fundaci\'on Antorchas and Balseiro.

\end{document}